\documentclass[prl,aps,twocolumn,showpacs,home,floats]{revtex4}

\usepackage{graphics}
\begin{document}

\title{{\rm\small\hfill (Phys. Rev. Lett., accepted)}\\
The steady-state of heterogeneous catalysis,\\ 
studied by first-principles statistical mechanics}

\author{Karsten Reuter$^{1,2}$}
\author{Daan Frenkel$^{2}$}
\author{Matthias Scheffler$^{1}$}

\affiliation{$^1$ Fritz-Haber-Institut der Max-Planck-Gesellschaft, Faradayweg 4-6, D-14195 Berlin, Germany}

\affiliation{$^2$ FOM Institute AMOLF, Kruislaan 407, SJ1098 Amsterdam, The Netherlands}

\begin{abstract}
The turn-over frequency of the catalytic oxidation of CO at
RuO$_2$(110) was calculated as function of temperature and 
partial pressures using {\em ab initio} statistical mechanics. 
The underlying energetics of the gas-phase molecules, dissociation, 
adsorption, surface diffusion, surface chemical reactions, and 
desorption were obtained by all-electron density-functional theory. 
The resulting CO$_2$ formation rate [in the full ($T, p_{\rm CO}, 
p_{\rm O_2}$)-space], the movies displaying the atomic motion and 
reactions over times scales from picoseconds to seconds, and the 
statistical analyses provide insights into the concerted actions 
ruling heterogeneous catalysis and open thermodynamic systems in 
general.
\end{abstract} 

\received{20 January 2004}


\maketitle

Under realistic conditions materials surfaces are in contact with a
rich environment \cite{Stampfl-SS500-2002}. Often, the resulting 
surface composition and surface-actuated material function are 
determined by equilibrium thermodynamics. A first-principles 
description is then possible with the ``{\em ab initio} atomistic 
thermodynamics'' approach, that has been successfully applied to 
various systems (see e.g. Refs. \cite{Stampfl-SS500-2002,
Reuter-PRB-2002,Reuter-PRL/B-2003,Michaelides-2003,Norskov-2003} 
and references therein). However, under many conditions 
equilibrium thermodynamics does not provide the appropriate
description, and heterogeneous catalysis is a particularly interesting 
and important example. Here one is dealing with an open system, i.e., 
a supply of gases or liquids comes into contact with a solid surface,
where a chemical reaction produces a new substance that is then 
transported away. The entire concert of the various underlying, 
interlinked atomistic processes is in this case determined by kinetics. 
Still, the temperature and partial pressures of the reactants must be set 
such that the system runs under {\em steady-state} conditions. Only
then the catalyst is not getting destroyed, but is stably enhancing
the rate of the desired chemical reaction. An {\em ab initio}
description of the full steady-state situation of heterogeneous
catalysis has not been achieved so far, and a microscopic
understanding of the competing and concerting actions of the various 
atomistic processes is lacking. 

The present paper describes the ``{\em ab initio} statistical
mechanics'' methodology appropriate for an open thermodynamic system 
with a continuous conversion of chemicals $A$ and $B$ into $C$, using
the oxidation of CO at a RuO$_2$ model catalyst as example. We employ
density-functional theory (DFT) to obtain the energetics of all 
relevant processes, as there are: motion of the gas-phase molecules, 
dissociation, adsorption, surface diffusion, surface chemical
reactions, and desorption. These calculations use the all electron,
full-potential linear augmented plane wave (FP-LAPW) approach 
\cite{wien,basisset}. The only notable approximations are the generalized 
gradient approximation (GGA) for the exchange-correlation functional 
\cite{PBE}, and the assumption that transition-state theory (TST) 
\cite{Ruggerone-1997} is applicable. In fact, both approximations are well 
justified for the present study, and we will particularly address the GGA below 
when analyzing the results. A combination of DFT with TST, and subsequently
solving the statistical mechanics problem by the kinetic Monte Carlo (kMC) 
approach \cite{Ruggerone-1997,Landau-2000}, has been employed before 
(see, e.g. Refs. \cite{Ruggerone-1997,Ovesson-1999,Kratzer-2002,Neurock-2000}).
In distinction to (by now standard) ``empirical kMC'' calculations, 
that use just a few {\em effective parameters}, which have only limited 
(if any) microscopic meaning, these ``{\em ab initio} kMC'' calculations 
include an extended set of elementary processes with full and direct 
physical meaning. And describing an open system with a continuous
conversion of chemicals $A$ and $B$ into $C$ implies additional 
complexity. For the example discussed below (the catalytic oxidation 
of CO) we will solve the statistical mechanics problem, gaining 
microscopic insight into the full dynamics from picoseconds to 
microseconds and even to seconds. The reported results demonstrate 
the new quality of and the novel insights gained by such description.
We will also compare the results to those obtained by ``constrained 
thermodynamics'' \cite{Reuter-PRL/B-2003}. Although (as expected) 
noticeable differences occur for certain environmental conditions, we 
will confirm that the much simpler ``constrained thermodynamics'' 
approach can provide important guidance on where in $(T,p)$-space 
high reaction rates are likely to be expected, and when results
obtained for certain $(T,p)$ conditions can be extrapolated to others. 

The now obtained turn-over-frequencies (TOFs) of the catalytic oxidation 
of CO are in unexpected and unprecedented agreement with experimental 
results of Peden and Goodman \cite{Peden-1986} and Wang {\em et al.}
\cite{Wang-2002}.  This is found to be a consequence of the fact that
the high reaction conditions are not just ruled by a singular chemical
reaction pathway. Instead, for the steady-state and high TOF conditions 
it is necessary that various processes play together in a most efficient 
manner. For the present example one crucial point is to realize an
optimum disordered and dynamic ``phase'' at the surface. Modest 
errors due to the approximate treatment of the exchange-correlation
functional (e.g. the GGA) are then not crucial, as long as the trends 
of the energetics of the various interplaying processes are described 
correctly. In this sense, the frequently requested {\em chemical
accuracy} for the description of individual processes appears to be a 
misleading concept. At least for the present system, a careful 
combination of DFT and statistical mechanics is more important. 

\begin{figure}[b!]
\scalebox{0.26}{\includegraphics{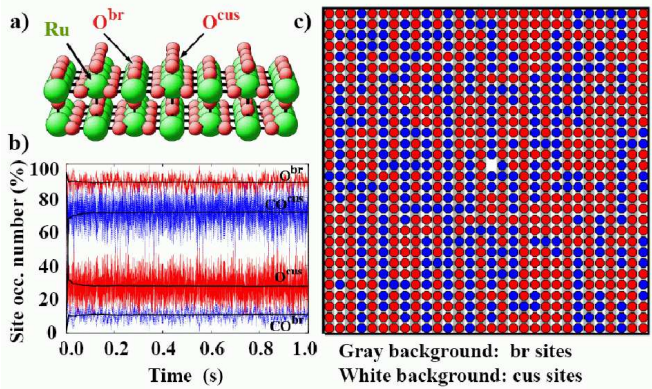}}
\caption{\label{snapshots}
a) Perspective view of the RuO${}_2$(110) surface, illustrating the two 
prominent adsorption sites in the rectangular surface unit cell. These 
sites are labeled as br (bridge) and cus (coordinatively unsaturated) site,
and both are occupied in the example with oxygen atoms. This is the ``high
$p_{\rm O_2}$ termination'' \cite{Reuter-PRB-2002,Reuter-PRL/B-2003}.
b) Time evolution of the site-occupation numbers from the
thermodynamic surface termination shown in a) to the steady-state 
of catalysis [cf. snapshot c)]. The temperature is $T= 600$\,K 
and the CO and O$_2$ partial pressures are that of the optimum
reaction conditions: $p_{\rm CO} = 20$\,atm, $p_{\rm O_2} = 1$\,atm.
c) Snapshot (top view of the RuO$_2$(110) surface) from movies
displaying the dynamics of the surface under realistic catalytic
conditions. Though the full atomic structure is considered in the 
calculations, for clarity we have marked here the substrate bridge 
sites only by gray stripes and the cus sites by white stripes. Oxygen 
adatoms are drawn as red circles and adsorbed CO molecules as blue 
circles. Movies for various pressure conditions for runtimes of
500 ns, as well as 1 s can be found at 
${\rm http:} \slash \slash$w3.rz-berlin.mpg.de$\slash\!\!\sim
$reuter$\slash$movies$\slash$movies.html.}
\end{figure}

\begin{table}
\caption{\label{table1}
DFT binding energies, $E_b$, for CO and O (with respect to (1/2)O$_2$) 
at br and cus sites (cf. Fig. \ref{snapshots}a), diffusion energy
barriers, $\Delta E^{\rm b}_{\rm diff}$, to neighboring br and cus 
sites, and reaction energy barriers, $\Delta E^{\rm b}_{\rm reac}$, of 
neighboring species at br and cus sites. All values are in eV.}
\begin{ruledtabular}
\begin{tabular}{l|c@{\quad}|cc@{\quad}|cc}
&  $E_b$ & \multicolumn{2}{c|}{$\Delta E^{\rm b}_{\rm diff}$} &    
\multicolumn{2}{c}{$\Delta E^{\rm b}_{\rm reac}$} \\
&        & to br & to cus & with CO$^{\rm br}$ & 
with CO$^{\rm cus}$ \\ \hline
CO$^{\rm br}$  & -1.6 & 0.6 & 1.6 & - & - \\
CO$^{\rm cus}$ & -1.3 & 1.3 & 1.7 & - & - \\
\hline
O$^{\rm br}$   & -2.3 & 0.7 & 2.3 & 1.5 & 1.2 \\
O$^{\rm cus}$  & -1.0 & 1.0 & 1.6 & 0.8 & 0.9 \\ 
\end{tabular}
\end{ruledtabular}
\end{table}

The methodology will be used to study CO oxidation over RuO${}_2$(110) 
(the surface is sketched in Fig. \ref{snapshots}a), as this system has 
recently received considerable attention as a highly active model 
catalyst (see Ref. \cite{Wang-2002} and references therein). In fact, 
this was previously called the Ru catalyst, but recent experimental
and theoretical work has shown that at realistic O$_2$ pressure the 
Ru(0001) surface is transformed into an epitaxial RuO$_2$(110) film 
(see Ref. \cite{Reuter-CPL-2002} and references therein). It is by 
now also established that this RuO$_2$(110) surface actuates the 
catalytic reaction, and, although domain boundaries and steps are 
present, their influence is not significant \cite{Jacobi-Ertl}. 
The surface unit cell is rectangular and contains two adsorption 
sites \cite{Reuter-PRL/B-2003}: bridge (br) and coordinatively 
unsaturated (cus) sites (cf. Fig. \ref{snapshots}a). Either of them can 
be empty or occupied by O or CO, and adsorbate diffusion can go br-to-br, 
br-to-cus, cus-to-cus, or cus-to-br. Since this comprises the possibility
of missing O$^{\rm br}$ atoms, we note that our treatment implicitly
includes the effect of O surface vacancies. A total of 26 
different elementary processes are possible on this lattice, and all 
were carefully analyzed by DFT to obtain their pathways and
energy barriers \cite{Reuter-PRL/B-2003,long-version}. Table \ref{table1} 
summarizes the adsorption energies and the diffusion and reaction 
energy barriers used for the kMC study, as obtained from DFT
calculations with a $(1 \times 2)$ surface unit-cell. In a systematic 
study of CO and O (co)adsorption at various coverages we found 
adsorbate-adsorbate interaction to be always smaller than 150\,meV. 
Thus, lateral interactions in this system are small (compared to 
the other energies), and will therefore be neglected.

CO adsorption into vacant cus or bridge sites is non-dissociative, while 
oxygen adsorption is dissociative and requires two vacant neighboring
sites, i.e. a br-br, cus-cus, or br-cus pair. The adsorption rate per
free site is given by the local sticking coefficient, $\tilde S$, 
and the kinetic impingement: $\Gamma_{\rm ad} = {\tilde S} \, p /
\sqrt{2\pi m k_{\rm B} T}$. Here $m$ is the mass of the gas-phase 
molecule, and $k_{\rm B}$ is the Boltzmann constant. The rate of the 
time reversed process (desorption) then follows from the relation:
$\Gamma_{\rm des} / \Gamma_{\rm ad} = {\rm exp} \left( (F_{b} - \mu)/
(k_{\rm B}T) \right)$, where $F_{b}$ is the free energy of the 
adsorbed species (approximated by $E_b$), and $\mu(T, p)$ is the 
chemical potential of the gas-phase molecule \cite{Reuter-PRB-2002}.
The $\tilde{S}(T)$ are thus obtained from the calculated total energy surfaces 
of desorption together with detailed balance. The only uncertainty 
here arises from the vibrational properties of the transition state 
which translates into uncertainties in the desorption rate by a factor
of 10 (at most 100). For the diffusion processes we use a 
prefactor of $10^{12}$\,Hz, which has an uncertainty of a factor of 10. 
We carefully checked that these uncertainties do not affect our below 
reported results and conclusions.
Details of the employed new methodology and of the various test calculations 
will be published elsewhere \cite{long-version}. Detailed balance 
of the scenario was carefully checked by confirming that the earlier results obtained by ``atomistic thermodynamics'' \cite{Reuter-PRL/B-2003} are exactly 
reproduced by the present statistical mechanics treatment, if surface reactions are not allowed to occur.

Kinetic Monte Carlo runs were performed for about 1000 different 
$(T,p_{\rm CO},p_{\rm O_2})$ conditions covering the temperature 
range $300\,{\rm K} < T < 800\,{\rm K}$ and partial pressures from 
$10^{-10}$ to 10$^{3}$ atmospheres. Several calculations were
done on a system with $(30 \times 30)$  surface sites (450 bridge 
plus 450 cus sites), but the vast amount of calculations was done 
for a 400 sites system. The results for both system sizes were
identical. The kMC simulations were run until steady-state is 
reached (cf. Fig. \ref{snapshots}b). Then the movies were recorded 
(cf. Fig. \ref{snapshots}c) and the statistical analyses of the 
frequencies of the various elementary processes performed. The 
latter also gives the total TOFs of the CO$_2$ formation.

The results show that catalytic reaction conditions are only
established when both CO and O$_2$ partial pressures exceed certain 
values below which the surface is in a thermodynamic equilibrium
phase of RuO${}_2$(110), characteristic for low oxygen pressures
and routinely observed under ultra-high vacuum (UHV) conditions:
All bridge sites are occupied by oxygen atoms and all cus sites are 
empty. For higher pressures three surface conditions are worth 
mentioning, which are e.g. obtained for $T = 600$\,K, $p_{\rm O_2} 
= 1$\,atm, and $p_{\rm CO} = 10^{-2}, 20,$ and $10^{3}$\,atm.

$(i)$ At the low $p_{\rm CO}$, all bridge and all cus sites are 
covered with oxygen atoms (cf. Fig. \ref{snapshots}a). This is 
essentially the thermodynamic high $p_{\rm O_2}$ phase 
\cite{Reuter-PRB-2002,Reuter-PRL/B-2003}. There is a noticeable
desorption/adsorption dynamics, i.e., about every 40\,$\mu$s two of 
the 900 O adatoms of the $(30 \times 30)$ simulation cell desorb 
(mainly from cus sites) as O$_2$. The resulting vacancies are then 
rapidly filled again (within 1\,ns), most of the time with oxygen. 
Only rarely CO adsorbs, and even if it does, it rather desorbs again
than initiating a reaction. The overall CO${}_2$ formation rate under
these conditions is with $0.9\!\cdot\!10^{12}$ cm$^{-2}$\,s$^{-1}$ very 
low. 

$(ii)$ For $p_{\rm CO} = 20$\,atm the situation corresponds to 
what was previously suggested to be that of high chemical 
activity \cite{Reuter-PRL/B-2003}, and this suggestion is now 
confirmed. The time to reach steady-state is remarkably long, in particular when
compared to the pico second timescale of the underlying atomistic 
processes (cf. Fig. \ref{snapshots}b). Finally, an interesting mix 
of CO molecules and O atoms at bridge and cus sites is established 
(a typical snapshot is shown in Fig. \ref{snapshots}c), 
with average occupation numbers $N_{\rm CO}^{\rm br} = 0.11$, 
$N_{\rm CO}^{\rm cus} = 0.70$, $N_{\rm O}^{\rm br} = 0.89$, and 
$N_{\rm O}^{\rm cus} = 0.29$. The dynamics of this surface is extremely 
fast. It is mainly due to CO desorption and adsorption. The rate of 
CO$_2$ formation is a factor of 0.0004 lower, but truly significant, namely 
$4.6 \cdot 10^{18}\,{\rm cm}^{-2}\,{\rm s}^{-1}$. It is dominated by 
the reaction CO$^{\rm cus}+ {\rm O}^{\rm cus} \longrightarrow {\rm CO}_2$.
The CO$^{\rm br} + {\rm O}^{\rm cus} \longrightarrow {\rm CO}_2$ and
CO$^{\rm cus}+ {\rm O}^{\rm br}  \longrightarrow {\rm CO}_2$ reactions
also contribute, though their rates are by a factor of 0.30 and 0.01 
lower than the first one. The CO$^{\rm br} + {\rm O}^{\rm br} 
\longrightarrow {\rm CO}_2$ reaction is insignificant. This result is 
different from what one would expect from the reaction energy barriers 
(cf. table \ref{table1}), which would give the lead to CO$^{\rm br}+ 
{\rm O}^{\rm cus}$, demonstrating the importance of the proper
mix of surface-site occupations.

$(iii)$ When the CO pressure is increased further, to $p_{\rm CO} =
10^3$\,atm, the surface becomes fully covered with CO. This result is 
at variance with the ``constrained thermodynamics'' study in which 
the CO$_2$ formation was forbidden. Now we find that catalysis is 
practically poisoned by adsorbed CO. The encountered situation is in 
fact close to that where the RuO$_2$ catalyst will be reduced by CO 
to the pure Ru metal.

\begin{figure}[b!]
\scalebox{0.38}{\includegraphics{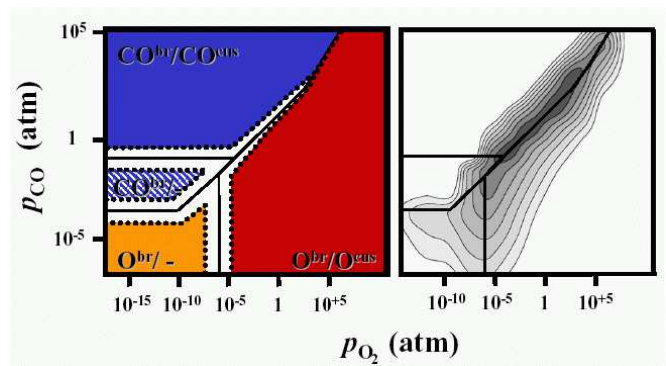}}
\caption{\label{TOF}
a) Steady-state surface structures obtained by {\em ab initio} kMC 
calculations at $T = 600$\,K and various CO and O$_2$ partial 
pressures. In all non-white areas, the average site occupation
is dominated ($> 90$\,\%) by one species, i.e. either O, CO or 
empty sites ($-$). b) Map of the corresponding turn-over frequencies 
(TOFs) in ${\rm cm}^{-2}{\rm s}^{-1}$: White areas have a TOF\,$<
10^{11} {\rm cm}^{-2}{\rm s}^{-1}$, and each increasing gray level 
represents one order of magnitude higher activity. Thus, the darkest
region corresponds to a TOF higher than $10^{17}\,{\rm cm}^{-2}{\rm s}^{-1}$.}
\end{figure}

\begin{figure}[t!]
\scalebox{0.38}{\includegraphics{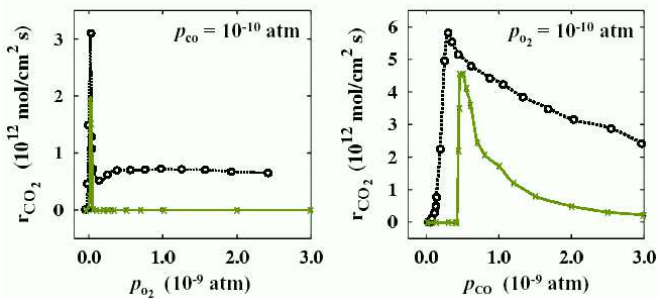}}
\caption{\label{Jacobi}
Rate of CO$_2$ formation at $T= 350$\,K. The experimental steady-state
results of Wang {\em et al.} \cite{Wang-2002} are presented as dotted lines, 
and the theoretical results are shown as green, solid lines. Rates
given as function of $p_{\rm O_2}$ at $p_{\rm CO}= 10^{-10}$\, atm (left), and
as function of $p_{\rm CO}$ at $p_{\rm O_2}= 10^{-10}$\, atm (right).}
\end{figure}

Figure \ref{TOF} summarizes the results by showing the various 
steady-state surface structures as well as a map of the TOFs in 
($p_{\rm CO}$, $p_{\rm O_2}$)-space. The highest activity is found 
to be in very good agreement with the early experimental results by 
Peden and Goodman \cite{Peden-1986}, and occurs whenever the
environmental conditions lead to the dynamic coexistence ``phase'' 
described above under $(ii)$. Recently, Wang {\em et al.} \cite{Wang-2002}
have also measured the TOFs for the RuO$_2$(110) surface at
$T =350$\,K for various pressures, and in Fig. \ref{Jacobi} we compare 
their results with ours. The agreement is again far better than what one 
would have expected: The theoretical and experimental TOF values at the 
optimum pressures are practically identical, and also the optimum
pressure conditions (when the ``dynamic phase'' is realized) agree 
very well.
Because the errors due to the GGA are in the range of 0.1-0.3\,eV, this 
good agreement between theory and experiment, seen in Fig. \ref{Jacobi}, 
may seem fortuitous. However, it is worth noting that the position of
the optimum catalytic efficiency in $(T,p)$-space and to some extent also
the value of the TOF are not determined by the energetics of a singular 
process alone, but by the action of many players. Apparently, 
the DFT-GGA calculations describe the differences between the various 
surface processes better than the individual absolute values. In this 
respect it is important to realize that a combination of different 
calculations (employing different approximations) or of theory and 
experiment could have spoiled the description. The consistent treatment 
of all participating processes implies that the {\em optimum mix} of O 
and CO at the surface is described well: The abundance of  
CO$^{\rm cus}$-O$^{\rm cus}$ nearest neighbor pairs (as well as 
CO$^{\rm br}$-O$^{\rm cus}$ nearest neighbor pairs) is apparently 
playing a role of similar importance as the energy barriers. Particularly 
at the optimum TOF conditions there is also an effective compensation, e.g.
when too high adsorption energies result in enhanced adsorption, but
also in reduced reaction barriers etc. Away from the optimum TOF 
conditions such compensation effects become less effective, and modest 
differences between the theoretical and experimental results arise. 
Here DFT-GGA errors are more influential, and for the experimental 
data we expect that contributions from surface imperfections may 
play a bigger role.

In summary, we computed the surface kinetics of CO oxidation catalysis 
at RuO$_2$(110). The TOFs are presented in $(T,p_{\rm CO},p_{\rm
O_2}$)-space, clearly identifying a narrow region of highest catalytic 
activity. In this region kinetics builds an adsorbate composition 
that is not found anywhere in the thermodynamic surface phase diagram.
The statistical analysis of the surface dynamics and of the various 
processes reveals several surprising results. For example, the 
chemical reaction with the most favorable energy barrier happens 
a factor of 0.30 less frequently than the energetically second 
favorable reaction.

The results also clarify how and when a bridging of the pressure gap 
between UHV studies and realistic pressure conditions is possible. The 
considered system has in fact two important components to this issue: 
At first, an O$_2$-rich environment changes the material from Ru to 
RuO$_2$. Thus, earlier high pressure studies on the Ru catalyst were 
actually looking at RuO$_2$. Second, after RuO$_2$ has been formed it 
is important to know how to set the pressure conditions correctly, as 
also on RuO$_2$(110) there is a low-pressure surface phase (the orange 
region in Fig. \ref{TOF}a), that has little in common with the
catalytically active situation. The obtained agreement between the 
theoretical results and experimental data confirms furthermore (cf. 
also Ref. \cite{Wang-2002}) that CO$_2$ is primarily formed from 
{\em adsorbed} CO and O, and that the metal-oxide, once it was created, 
does not play an active role, i.e., there is no indication of significant 
bulk diffusion. Thus, the catalysis is explained in terms of a 
Langmuir-Hinschlwood mechanism. Some contributions via the Eley-Rideal 
mechanism (e.g. scattering of gas-phase CO at O$^{\rm cus}$) may 
play a minor role, but this process requires further calculations and 
experiments. With the advent of first-principles TOF-maps obtained from 
DFT calculations (cf. Fig. \ref{TOF}), a more detailed data base is 
in general becoming available for comparison with experiments, which will
eventually advance the microscopic understanding of catalysis.

\end{document}